\documentclass[aps]{revtex4}
\newcommand{\Cnl}{C_{\rm nl}}
\newcommand{\D}{\mathbf{\mathcal{D}}}
\newcommand{\X}{\mathbf{X}}
\newcommand{\etal}{\emph{et al.}}
\newcommand{\Hsp}{H_{\mathrm{sp}}}
\newcommand{\Vad}{V_{\mathrm{ad}}}
\usepackage{graphicx}
\begin{document}

\title{Calculation of the microcanonical temperature 
for the classical Bose field}
\author{M.~J.~Davis}
\email{mdavis@physics.uq.edu.au}
\affiliation{ARC Centre of Excellence for Quantum Atom Optics, 
School of Physical Sciences, University of Queensland, Brisbane, QLD 
4072, Australia}
\author{P.~B.~Blakie}
\affiliation{Department of Physics, University of Otago, P.O. Box 56, Dunedin,
New Zealand.}


\begin{abstract}

The ergodic hypothesis asserts that a classical mechanical system will in time
visit every available configuration in phase space.  Thus, for an ergodic
system, an ensemble average of a thermodynamic quantity can equally well be
calculated by a time average over a sufficiently long period of dynamical
evolution.  In this paper we describe in detail how to calculate the temperature
and chemical potential from the dynamics of a microcanonical classical field,
using the particular example of the classical modes of a Bose-condensed gas. 
The accurate determination of these thermodynamics quantities is essential in
measuring the shift of the critical temperature of a Bose gas due to
non-perturbative many-body effects.

\end{abstract}
\maketitle

\section{Introduction}

The shift in critical temperature of the homogeneous Bose gas has been the
subject of numerous investigations over the past fifty years.  As the density of
this idealized system is constant, the shift due to the mean-field
is zero, and the first order shift is due to
long-wavelength critical fluctuations.  The first estimates were due to
 Lee and Yang \cite{Lee1957a,Lee1958a}, who gave two different results for the
dependence on the s-wave scattering length $a$.
  In 1999 Baym \emph{et al.}\
\cite{Baym1999a} determined that the result should be 
\begin{equation}
\Delta T_c/T_c^0
= c a n^{1/3},
\end{equation}
where $n$ is the particle number density, and $c$ is a constant of order unity.  Several authors have
attempted to calculate this constant, and a wide range of results have been
obtained, as summarised in Fig.~1 of \cite{Arnold2001c}.  However recent Monte
Carlo calculations have apparently settled this matter, giving a combined result
of $c \approx 1.31 \pm 0.02$ \cite{Arnold2001c,Kashurnikov2001a}.  A useful
summary and discussion of this topic is provided by 
Andersen \cite{Andersen2004a} and Holzmann \emph{et al}.~\cite{Holzmann2004a}.

Previously we have performed numerical simulations of an equation known
as the Projected Gross-Pitaevskii equation (PGPE), which can be used to
represent the highly occupied modes of Bose condensed gases at finite
temperature \cite{Svistunov1991,Kagan1992,Kagan1994,Kagan1997,Davis2001a}. This equation is observed to evolve randomised
initial conditions to equilibrium for which it is possible to measure a
temperature \cite{Davis2001b,Davis2002a, Blakie2005a}.   The PGPE is
non-perturbative, and hence includes the effect of critical fluctuations. 
The only approximation made is that the modes of the gas
are sufficiently highly occupied as to be well described by a classical
rather than a quantum field.  The occupation condition is that 
 mode $k$ must satisfy $\langle N_k\rangle \gg 1$; however for practical
simulations we may choose, for example, $\langle N_k\rangle \ge 5$
\cite{Blakie2005b}.
This method is suitable for investigating many problems 
of current interest in ultra-cold Bose gases, 
such as the shift in critical temperature due to non-perturbative effects
\cite{Davis2005a}.

The PGPE describes a microcanonical system, with the classical field restricted
to a finite number of modes for which the occupation number condition is met. In
order to study the problem of the shift in $T_c$ it is necessary to
accurately determine thermodynamic quantities defined as derivatives of
the entropy such as the temperature and chemical potential.
In 1997 Rugh developed a new
approach to statistical mechanics where he derived an expression from the
Hamiltonian of a system for which the ensemble average gives the temperature
within the microcanonical ensemble \cite{Rugh1997a}. 
However, if the system is known to be ergodic then the equilibrium temperature
can be determined from the system dynamics over a sufficiently long period of 
time.  

We have applied an extension of Rugh's method to the PGPE Hamiltonian, and the
appropriate expression to determine the temperature is given as Eq.~(22) of
\cite{Davis2003a}.  This method was found to agree with the less rigorous
methods described in \cite{Davis2001b,Davis2002a}.  In \cite{Davis2003a} we
made use of this method to calculate the shift in the critical temperature of
the homogeneous Bose gas.  Despite the calculation being performed with limited
statistics and suffering from finite size effects, it gave a result of $c=
1.3\pm 0.4$ in agreement with the Monte Carlo results 
\cite{Arnold2001c,Kashurnikov2001a}.  In \cite{Davis2005a} we applied this
method to the experiment of Gerbier \emph{et al.} \cite{Gerbier2004a}  who
measured the shift in critical temperature of a trapped Bose gas, and found good
agreement with experiment.
In this paper we give the details of our
implementation of Rugh's method for a general mode basis for the PGPE.

\section{Background}

\subsection{Formalism of Rugh}

We consider a classical system with $M$ independent modes. 
 The  Hamiltonian can
be written as $H = H(\mathbf{\Gamma})$, where $\mathbf{\Gamma} = \{\Gamma_i\} =
\{Q_i,P_i\}$ is a vector of length $2M$ consisting of the canonical position
and momentum co-ordinates.  We define the gradient
operator $\nabla$ in terms of its components $\nabla_i = \partial/\partial\Gamma_i$.
In the notation of Rugh \cite{Rugh2001a}, the Hamiltonian $H$ may have a number of
independent first integrals, labelled by $F = F_1, \ldots,F_m,$ 
that are invariant under the dynamics of $H$.  A particular macro-state of
such a system can be specified by the values of the conserved quantities,
labelled as $H = E, F_i = I_i$.

The usual expression for the temperature of a system in the microcanonical ensemble is given by
\begin{equation}
\frac{1}{k_B T} 
= 
\left(\frac{\partial S}{\partial E}\right)_{F_i},
\end{equation}
where all other constants of motion are held fixed, and where the entropy 
can be written
\begin{equation}
e^{S/k_B} =
 \int d\mathbf{\Gamma} \;\delta[E - H(\mathbf{\Gamma})]
\; \prod_i \delta[{I_i -
 F_i(\mathbf{\Gamma})}].
\end{equation}
Using Rugh's methods, the temperature of the
system can be written as
\begin{equation}
\frac{1}{k_B T} 
= 
\bigg\langle \mathbf{\D} \cdot \mathbf{X}_T(\mathbf{\Gamma}) \bigg\rangle,
\label{eqn:temp_eqn}
\end{equation}
where the angle brackets correspond to an ensemble average, and the components of the vector operator $\D$ are
\begin{equation}
\D_i = e_i \frac{\partial}{\partial \Gamma_i},
\end{equation}
where $e_i$ can be chosen to be any scalar value, including zero.
The vector field $\mathbf{X}_T$ can also be chosen freely within the constraints
\begin{eqnarray}
\D H \cdot \mathbf{X}_T =1,
\quad \D F_i \cdot \mathbf{X}_T = 0, \quad 1\le i \le m.
\label{eqn:X_conditions}
\end{eqnarray}
Geometrically this means that the vector field $\mathbf{X}_T$ has a non-zero
component transverse to the $H = E$ energy surface, and is parallel to the surfaces
$F_i = I_i$. The expectation value in  
Eq.~(\ref{eqn:temp_eqn}) is over all possible states in the microcanonical ensemble;
however if the ergodic theorem is applicable then it can equally well be interpreted
as a time-average.  For further details on the origin of this expression we refer
the reader to  Rugh's original papers \cite{Rugh1997a,Rugh1998a,Rugh2001a}, as well as
derivations  found in Giardin\`{a} and Levi \cite{Giardina1998a}, 
Jepps \etal\
\cite{Jepps2000a} and Rickayzen and Powles \cite{Rickayzen2001a}.

\subsection{Dimensionless Hamiltonian}
The classical Hamiltonian for the dilute Bose gas in dimensionless form is
\begin{equation}
H  = \int d^3 \mathbf{x} \left[  
\psi^*(\mathbf{x}) \Hsp \psi(\mathbf{x}) +
\Vad(\mathbf{x})|\psi(\mathbf{x})|^2 +
\frac{\Cnl}{2}
|\psi(\mathbf{x})|^4\right],
\label{eqn:qH}
\end{equation}
where $H = \tilde{H}/(N_C \epsilon_L) $, 
$N_C$ is the number of particles in the system, $\mathbf{x} = 
\mathbf{\tilde{x}}/L$, 
$L$ is the unit of length, 
$\epsilon_L = \hbar^2/(2mL^2)$ is the unit of energy, and $m$ is the mass of the
particles. 
The dimensionless classical Bose field operator
$\psi(\mathbf{x})$ 
is here normalized to one,
 $\int d^3 
  \mathbf{x} |\psi(\mathbf{x})|^2=1$,
and $\Cnl$ is the nonlinear constant defined as
\begin{equation}
\Cnl = \frac{N_C U_0}{\epsilon_L L^3} = \frac{8\pi a N_C}{L},
\end{equation}
where $U_0 = 4\pi \hbar^2 a / m $.
$\Hsp$ is the single particle Hamiltonian with eigenbasis 
$\Hsp \phi_k(\mathbf{x}) = \epsilon_k\phi_k(\mathbf{x})$, and
$\Vad(\mathbf{x})$ is the dimensionless version of any
 potential \emph{additional} to that included in the single-particle
 Hamiltonian.

We restrict our field $\psi(\mathbf{x})$ to be formed from the modes of the classical
region $C$ defined by a high energy cutoff in the single-particle basis,
 such that we can write
\begin{equation}
\psi(\mathbf{x}) = \sum_{k \in C} c_k \phi_k(\mathbf{x}).
\end{equation}
The equation of motion for this restricted system is the Projected
Gross-Pitaevskii equation
\begin{eqnarray}
i\frac{\partial\psi(\mathbf{x})}{\partial \tau}  &=& 
\Hsp \psi
 +
\mathcal{P}\{ \Vad(\mathbf{x})\psi(\mathbf{x}) +
 C_{\rm nl}|\psi(\mathbf{x})|^{2}\psi(\mathbf{x})\}.
\label{eq:GPE1}
\end{eqnarray}
 The projection operator $\mathcal{P}\{F\}$ projects function $F$ 
onto the classical modes of the single particle Hamiltonian $\Hsp$ via
\begin{equation}
{\mathcal{P}}\{F(\mathbf{x}) \}
= \sum_{k \in C}\phi_k({\bf x}) \int d^3{\bf x}' \;\phi_k^*({\bf x}')
F({\bf x}').
\end{equation}
It eliminates collisional processes which would cause the 
scattering of particles into higher energy single particle modes not 
represented by the classical field.

In the single-particle basis of $\Hsp$ the Hamiltonian (\ref{eqn:qH}) can be written
\begin{equation}
H = \sum_{n\in C} \epsilon_n c_n^* c_n
+ \sum_{mn\in C} V_{mn} c_m^* c_n
+ \frac{\Cnl}{2} \sum_{mnpq\in C} 
c_m^* c_n^* c_p c_q
\langle m n|p q\rangle,
\label{eqn:Hbasis}
\end{equation}
where we have defined the matrix elements 
\begin{eqnarray}
V_{mn} &=& \int d^3\mathbf{x}\, \phi^*_m(\mathbf{x})\Vad(\mathbf{x})
 \phi_n(\mathbf{x}),
\\ 
\langle m n | p q \rangle &=& \int d^3\mathbf{x} \,
\phi^*_m(\mathbf{x}) \phi^*_n(\mathbf{x})   
\phi_p(\mathbf{x})  \phi_q(\mathbf{x}).
\label{eqn:matrix_element}
\end{eqnarray}
We make use of the real, canonically-conjugate coordinates $Q_n$ and $P_n$
defined by 
\begin{equation}
Q_n = \frac{1}{\sqrt{2 \epsilon_n}}(c_n^* + c_n), \quad
P_n = i\sqrt{\frac{\epsilon_n}{2}}(c_n^* - c_n),
\label{def:XP}
\end{equation}
with the inverse transformation
\begin{equation}
c_n = \sqrt{\frac{\epsilon_n}{2}} Q_n + \frac{i}{\sqrt{2 \epsilon_n}} P_n, \quad
c_n^* = \sqrt{\frac{\epsilon_n}{2}} Q_n - \frac{i}{\sqrt{2 \epsilon_n}} P_n.
\end{equation}

\subsection{Choice of vector field $\mathbf{X}$}
\label{blergh}
In order to satisfy the conditions (\ref{eqn:X_conditions}) we can 
choose a vector field of the form
\begin{equation}
\mathbf{X}_T = a \D H + \sum_{i=1}^{m} b_i  \D F_i,
\label{eqn:X_expansion}
\end{equation}
where the $m+1$ coefficients $\{a,b_i\}$
are determined by the $m+1$ simultaneous equations in Eq.~(\ref{eqn:X_conditions}). 
Due to the freedom in the choice of 
the vector operator $\D$
 we can set any number of components of the length $2M$ vector $\mathbf{X}_T$ to zero.
This
turns out to be useful as the components corresponding to the momentum and
position variables can be different orders of magnitude.  
Two particular choices
we make use of later are $\X_{T,P}$ with $\D = \D_P = \{0,\partial/\partial P_i\}$
and $\X_{T,Q}$
with $\D = \D_Q= \{ \partial/\partial Q_i,\,0\}.$
These lead to two different calculations for the temperature that only agree
in general if the system is in thermal equilibrium.  This provides a 
useful check that the numerical simulations have indeed reached thermal
equilibrium, as well as giving two independent values for the temperature.

For the classical Bose gas Hamiltonian  (\ref{eqn:qH}) there can be several constants
of motion that must be taken into account.  These depend on the details of the
single-particle Hamiltonian $\Hsp$ --- examples include components of the angular
and/or linear momentum.   The effect of including these additional first
integrals in the definition of the vector field $\mathbf{X}_T$ is to account for
the energy that is associated with a conserved quantity and hence is unavailable
for thermalization. This ensures that only the appropriate free energy is used
to calculate the temperature. We conjecture that the same result can be achieved
by first transforming to a co-ordinate system  where the total angular and
linear momenta, etc, are all zero and therefore do not contribute to the energy
of the system.  Rugh demonstrated this explicitly in \cite{Rugh2001a} for a
system of particles with a conserved centre-of-mass motion, and our numerical
results support this conjecture.

An exception to this, however, is the conservation of normalization  $N = \sum_n
c^*_n c_n$. This must be considered explicitly because there is no co-ordinate
system in which it can be made to vanish.  The constraint on $N$ means that the
ground state of the system will, in general, have a finite energy that is not
accessible for thermalization.  We note that the effect of this constraint is in
general more complicated than a simple subtraction of the ground state energy
(which could be achieved by hand) and depends on the definition of the operator
$\D$ used to calculate the temperature, as shown below.

We need to choose a vector field
$\mathbf{X}_T$ which satisfies Eqs.~(\ref{eqn:X_conditions}) with $F_1 = N = \sum_n
c^*_n c_n$. The result is
\begin{equation}
\mathbf{X}_T = \frac{\D H - \lambda_N\D N}
{|\D H|^2 - \lambda_N\D N \cdot \D H},
\label{eqn:Xvec}
\end{equation}
where the parameter
\begin{equation}
\lambda_N = \frac{\D N \cdot \D H}{|\D N|^2},
\label{eqn:X_with_N}
\end{equation}
looks similar to a chemical potential. 
Substituting Eq.~(\ref{eqn:Xvec}) into Eq.~(\ref{eqn:temp_eqn}) we find that 
our full expression for the dimensionless temperature $\tilde{T} = k_B T/ N_C
\epsilon_L$ is
\begin{widetext}
\begin{eqnarray}
\frac{1}{\tilde{T}} = \left\langle
\frac{\D^2 H - \lambda_N \D^2N - \D\lambda_N\cdot\D N}
{|\D H|^2 - \lambda_N (\D H \cdot \D N)}
\right\rangle
-
\left\langle
\frac{ (\D H - \lambda_N\D N) \cdot [ \D |\D H|^2 - (\D H \cdot \D N) \D \lambda_N
 - \lambda_N \D (\D H \cdot \D N)]}
{[{|\D H|^2 - \lambda_N (\D H \cdot \D N)}]^2}
\right\rangle,
\label{eqn:T}
\end{eqnarray}
\end{widetext}
where
\begin{equation}
\D \lambda_N = \frac{\D (\D N \cdot \D H)}{|\D N|^2}
+ \frac{\D N \cdot \D H}{|\D N|^4} \D (|\D N|^2).
\label{eqn:Dl}
\end{equation}

The chemical potential of the system can  be determined in a similar manner
using a different choice of vector field.  In particular we wish to calculate 
\begin{equation}
\frac{\mu}{k_B T} 
= 
- \left(\frac{\partial S}{\partial N}\right)_{E} =
 \bigg\langle \mathbf{\D} \cdot \mathbf{X}_\mu(\mathbf{\Gamma}) \bigg\rangle,
\end{equation}
where the conditions on the vector field are
\begin{eqnarray}
\D H \cdot \mathbf{X}_\mu =0,
\quad \D N \cdot \mathbf{X}_\mu = 1.
\label{eqn:X_conditions2}
\end{eqnarray}
This results in expressions identical to the RHS of
Eqs.~(\ref{eqn:Xvec},\ref{eqn:X_with_N},\ref{eqn:T}) but with $H$ and $N$
interchanged, and so we will not write them out in full.

The technique outlined above has been described previously by one of us in 
Ref.~\cite{Davis2003a}, and in particular the result of Eq.~(\ref{eqn:T}) was given
in that work.  The numerical method has also been applied by the current
authors in Refs.~\cite{Blakie2005a,Davis2005a}.  The purpose of this paper is
to outline in detail the procedure for calculating the terms in
Eqs.~(\ref{eqn:T}) and (\ref{eqn:Dl}).  This is not a trivial exercise, and the
proliferation of terms can make it difficult to
avoid errors in both the analytical and numerical implementation. 
By discussing our approach to evaluating Eqs.~(\ref{eqn:T}) and 
(\ref{eqn:Dl}) we hope to facilitate  other researchers wanting to make 
use of the method.
  The basic numerical requirement is an
efficient and accurate numerical
transform from real space to basis space and vice versa.  This may be provided,
for example, by fast Fourier transforms using a basis of plane waves appropriate
for a homogeneous system \cite{Davis2001b,Davis2002a}, 
or an efficient quadrature for trapped systems \cite{Blakie2005a}.

\section{Formulae} 

In this section we give explicit analytic
expressions for all quantities required in the
evaluation of Eq.~(\ref{eqn:T}) and (\ref{eqn:Dl}). 
 As we are using the canonical
co-ordinates defined in Eq.~(\ref{def:XP}) it is easiest to evaluate the
derivatives of the Hamiltonian using the mode expansion of the Hamiltonian
(\ref{eqn:Hbasis}).  However, this leaves us with expressions involving inner
products of vector quantities with summations that would be
computationally expensive to evaluate directly for any sizeable basis.  We
show how to define auxiliary field functions that can be used to simplify
these terms for implementation with an efficient numerical transform.

\subsection{Derivatives}

We begin by writing out the necessary derivatives of the Hamiltonian using the
two choices of vector operator defined earlier, $\D_Q$ and $\D_P$. 
 In component
form the first derivatives 
of the Hamiltonian are
\begin{eqnarray}
(\D_Q H)_i &=& \epsilon_i^2 Q_i 
+ \sqrt{\frac{ \epsilon_i}{2}}\sum_n \left[V_{ni} c_n^* + V_{in} c_n\right]
+ \Cnl \sqrt{\frac{ \epsilon_i}{2}}
\sum_{pqm} \left[ c^*_p c^*_q c_m \langle pq|mi\rangle
+ c^*_m c_p c_q \langle mi|pq\rangle\right],
\label{DQH1}
\\
(\D_P H)_i &=& P_i 
+ \frac{i}{\sqrt{2 \epsilon_i}}\sum_n \left[V_{ni} c_n^* - V_{in} c_n\right]
+ i \frac{ \Cnl}{\sqrt{2 \epsilon_i}}
\sum_{pqm} \left[ c^*_p c^*_q c_m \langle pq|mi\rangle
- c^*_m c_p c_q \langle mi|pq\rangle\right],
\label{DPH1}
\end{eqnarray}
and the second derivatives are
\begin{eqnarray}
(\D_Q^2 H)_{ij} &=& \delta_{ij} \epsilon_i^2 + 
\frac{\sqrt{\epsilon_i \epsilon_j}}{2}[V_{ij} + V_{ji}]
+ \frac{\Cnl}{2} \sqrt{\epsilon_i \epsilon_j}
\sum_{pq} \left[ 4 c^*_p c_q \langle pi|qj\rangle
+ c_p c_q \langle ij|pq\rangle   + c_p^* c_q^* \langle pq|ij\rangle \right],
\label{DQH2}
\\
(\D_P^2 H)_{ij} &=& \delta_{ij} + 
\frac{1}{2\sqrt{\epsilon_i \epsilon_j}}[V_{ij} + V_{ji}]
+ \frac{\Cnl}{2\sqrt{\epsilon_i \epsilon_j}}
\sum_{pq} \left[ 4c^*_p c_q  \langle pi|qj\rangle
- c_p c_q \langle ij |pq\rangle - c_p^* c_q^* \langle pq|ij\rangle\right].
\label{DPH2}
\end{eqnarray}
We also need the derivatives of the other first integral of the Hamiltonian ---  the
the normalisation given by $N = \sum_{k \in C} |c_k|^2$.  These are
\begin{eqnarray}
(\D_Q N)_i = \epsilon_i Q_i,\qquad&&\qquad
(\D_P N)_i =\frac{P_i}{\epsilon_i},
\label{eqn:list1}
\\
(\D_Q^2 N)_{ij} =\epsilon_i \delta_{ij},\qquad&&\qquad
(\D_P^2 N)_{ij} =\frac{\delta_{ij}}{\epsilon_i}.
\end{eqnarray}

From Eq.~(\ref{eqn:T}) we can see that we need the terms
\begin{equation}
\D (|\D H|^2)_i = 2 \sum_j (\D H)_j (\D^2 H)_{ij},
\qquad \D (|\D N|^2)_i = 2 \sum_j (\D N)_j (\D^2 N)_{ij},
\end{equation}
\begin{equation}
\D (\D N \cdot \D H)_i =  
\sum_j\left[ (\D N)_j (\D^2 H)_{ij} + (\D H)_j (\D^2 N)_{ij}\right].
\label{eqn:list2}
\end{equation}
Some of these expressions are simple to calculate e.g.
\begin{eqnarray}
\D_Q (|\D_Q N|^2)_i &=& 2 \sum_j (\D_Q N)_j (\D_Q^2 N)_{ij}\\
&=&
2 \sum_j (\epsilon_j Q_j) (\epsilon_i \delta_{ij})
\;=\; 2\epsilon_i^2 Q_i,
\end{eqnarray}
\begin{eqnarray}
\D_P (|\D_P N|^2)_i &=& 2 \sum_j (\D_P N)_j (\D_P^2 N)_{ij}\\
&=&
2 \sum_j \left(\frac{P_i}{\epsilon_i}\right) \left(\frac{\delta_{ij}}{\epsilon_i}\right)
\;=\; 2\frac{ P_i}{\epsilon_i^2},
\end{eqnarray}
\begin{eqnarray}
\sum_j (\D_Q H)_j (\D_Q^2 N)_{ij}
&=& \sum_j (\D_Q H)_j \epsilon_i \delta_{ij} \;=\;
\epsilon_i (\D_Q H)_i,
\end{eqnarray}
\begin{eqnarray}
\sum_j (\D_P H)_j (\D_P^2 N)_{ij}
&=&\frac{\delta_{ij}}{\epsilon_i} \;=\;
\frac{(\D_P H)_i}{\epsilon_i}.
\end{eqnarray}
However, the remainder are more complicated.
To calculate them efficiently we make the following vector definitions
\begin{eqnarray}
a_j &=& \sqrt{\epsilon_j} (\D_Q H)_j,\qquad
b_j =  \frac{(\D_P H)_j}{\sqrt{\epsilon_j}},\\
f_j &=& \sqrt{\epsilon_j} (\D_Q N)_j,\qquad
g_j =  \frac{(\D_P N)_j}{\sqrt{\epsilon_j}},
\end{eqnarray}
with the corresponding auxiliary field functions
\begin{eqnarray}
A(\mathbf{x}) &=& \sum_j a_j \phi_j(\mathbf{x}),\qquad
B(\mathbf{x}) =  \sum_j b_j \phi_j(\mathbf{x}),
\label{eqn:aux1}\\
F(\mathbf{x}) &=& \sum_j f_j \phi_j(\mathbf{x}),\qquad
G(\mathbf{x}) =  \sum_j g_j \phi_j(\mathbf{x}).
\label{eqn:aux2}
\end{eqnarray}
Making use of these definitions we find
\begin{eqnarray}
\D_Q (|\D_Q H|^2)_i &=& 2 \sum_j (\D_Q H)_j (\D_Q^2 H)_{ij},
\nonumber\\
&=& 2 \sum_j \frac{a_j}{\sqrt{\epsilon_j}}\left(\delta_{ij} \epsilon_i^2 + 
\frac{\sqrt{\epsilon_i \epsilon_j}}{2}[V_{ij} + V_{ji}]
+ \frac{\Cnl}{2} \sqrt{\epsilon_i \epsilon_j}
\sum_{pq} \left[ 4 c^*_p c_q \langle pi|qj\rangle
+ c_p c_q \langle ij|pq\rangle   + c_p^* c_q^* \langle pq|ij\rangle
\right]\right),
\nonumber\\
&=& 2 a_i \epsilon_i^{3/2} + 
\sqrt{\epsilon_i}\sum_j a_j[V_{ij} + V_{ji}]
+ \Cnl \sqrt{\epsilon_i }
\sum_{pqj}a_j \left[ 4 c^*_p c_q \langle pi|qj\rangle
+ c_p c_q \langle ij|pq\rangle   + c_p^* c_q^* \langle pq|ij\rangle \right],
\label{DQDQH2}
\end{eqnarray}
\begin{eqnarray}
\D_P (|\D_P H|^2)_i &=& 2 \sum_j (\D_P H)_j (\D_P^2 H)_{ij},
\nonumber\\
&=&2\sum_j\sqrt{\epsilon_j}b_j
\left(\delta_{ij} + 
\frac{1}{2\sqrt{\epsilon_i \epsilon_j}}[V_{ij} + V_{ji}]
+ \frac{\Cnl}{2\sqrt{\epsilon_i \epsilon_j}}
\sum_{pq} \left[ 4c^*_p c_q  \langle pi|qj\rangle
- c_p c_q \langle ij |pq\rangle - c_p^* c_q^* \langle pq|ij\rangle\right ]
\right),\nonumber\\
&=&2\sqrt{\epsilon_i}b_i + \frac{1}{\sqrt{\epsilon_i}}
\sum_j b_j[V_{ij} + V_{ji}]
+ \frac{\Cnl}{\sqrt{\epsilon_i}}
\sum_{pqj}b_j \left[ 4c^*_p c_q  \langle pi|qj\rangle
- c_p c_q \langle ij |pq\rangle - c_p^* c_q^* \langle pq|ij\rangle\right ],
\label{DPDPH2}
\end{eqnarray}
and
\begin{eqnarray}
\sum_j (\D_Q N)_j (\D_Q^2 H)_{ij}
&=&
\sum_j\frac{f_j}{\sqrt{\epsilon_j}}
\left(\delta_{ij} \epsilon_i^2 + 
\frac{\sqrt{\epsilon_i \epsilon_j}}{2}[V_{ij} + V_{ji}]
+ \frac{\Cnl}{2} \sqrt{\epsilon_i \epsilon_j}
\sum_{pq} \left[ 4 c^*_p c_q \langle pi|qj\rangle
+ c_p c_q \langle ij|pq\rangle   + c_p^* c_q^* \langle pq|ij\rangle \right]
\right),\nonumber\\
&=&
 f_i \epsilon_i^{3/2} + 
\frac{\sqrt{\epsilon_i}}{2}\sum_j f_j[V_{ij} + V_{ji}]
+ \frac{\Cnl}{2} \sqrt{\epsilon_i }
\sum_{pqj}f_j \left[ 4 c^*_p c_q \langle pi|qj\rangle
+ c_p c_q \langle ij|pq\rangle   + c_p^* c_q^* \langle pq|ij\rangle \right],
\label{DQND2QH}
\end{eqnarray}

\begin{eqnarray}
\sum_j (\D_P N)_j (\D_P^2 H)_{ij}
&=&\sum_j \sqrt{\epsilon_j}g_j
\left(\delta_{ij} + 
\frac{1}{2\sqrt{\epsilon_i \epsilon_j}}[V_{ij} + V_{ji}]
+ \frac{\Cnl}{2\sqrt{\epsilon_i \epsilon_j}}
\sum_{pq} \left[ 4c^*_p c_q  \langle pi|qj\rangle
- c_p c_q \langle ij |pq\rangle - c_p^* c_q^* \langle pq|ij\rangle\right ]
\right),\nonumber\\
&=&\sqrt{\epsilon_i}g_i + \frac{1}{ 2 \sqrt{\epsilon_i}}
\sum_j g_j[V_{ij} + V_{ji}]
+ \frac{\Cnl}{2\sqrt{\epsilon_i}}
\sum_{pqj}g_j \left[ 4c^*_p c_q  \langle pi|qj\rangle
- c_p c_q \langle ij |pq\rangle - c_p^* c_q^* \langle pq|ij\rangle\right ].
\label{DPND2PH}
\end{eqnarray}

\subsection{Evaluation of terms with auxiliary fields}
The above expressions are  written in a form that they can be easily
calculated using efficient quadratures in real space.  We outline in detail
how to calculate all of these terms, and write them in the most
efficient form we have found.  We make the definitions
\begin{eqnarray}
\sum_n V_{in} c_n &=& 
\int d^3\mathbf{x}  \phi_i^*(\mathbf{x})
\Vad(\mathbf{x})\psi(\mathbf{x})\;\equiv\; W_i,
\\
\sum_n V_{ni} c_n^* &=& \int d^3\mathbf{x} \psi^*(\mathbf{x}) 
\Vad(\mathbf{x})
\phi_i(\mathbf{x})
\;\equiv\; W_i^*,
\end{eqnarray}
where we have assumed that the potential $\Vad(\mathbf{x})$ is real. 
We  can therefore 
write the appropriate parts of Eqs.~(\ref{DQH1}) and (\ref{DPH1}) as
\begin{eqnarray}
\sum_n \left[V_{ni} c_n^* + V_{in} c_n\right] &=& 2 \mbox{Re}(W_i),\\
\sum_n \left[V_{ni} c_n^* - V_{in} c_n\right] &=& -2 i \mbox{Im}(W_i).
\end{eqnarray}
The next set of terms are
\begin{eqnarray}
\sum_{pqm} c^*_m c_p c_q \langle mi|pq\rangle &=&
\int d^3\mathbf{x} \phi_i^*(\mathbf{x})|\psi(\mathbf{x})|^2 \psi(\mathbf{x}) 
\;\equiv\; K_i,\\
\sum_{pqm} c^*_p c^*_q c_m \langle pq|mi\rangle
&=&
\int d^3\mathbf{x} |\psi(\mathbf{x})|^2 \psi^*(\mathbf{x}) \phi_i(\mathbf{x})
\;\equiv\; K_i^*.
\end{eqnarray}
Hence, parts of Eqs.~(\ref{DQH1}) and (\ref{DPH1}) can be written
 \begin{eqnarray}
\sum_{pqm} \left[ c^*_p c^*_q c_m \langle pq|mi\rangle
+ c^*_m c_p c_q \langle mi|pq\rangle\right]
&=&
2 \mbox{Re}(K_i),\\
\sum_{pqm} \left[ c^*_p c^*_q c_m \langle pq|mi\rangle
- c^*_m c_p c_q \langle mi|pq\rangle\right]
&=& -2i \mbox{Im}(K_i).
\end{eqnarray}
For the terms involving the auxiliary fields we have expressions like
\begin{eqnarray}
\sum_{pqj}a_j  c^*_p c_q \langle pi|qj\rangle
& = &
\int d^3\mathbf{x} \phi_i^*(\mathbf{x})
|\psi(\mathbf{x})|^2 A(\mathbf{x}) \;\equiv\; (A_1)_i,
\label{eqn:A1}\\
\sum_{pqj}a_j  c_p c_q \langle ij|pq\rangle&=&
\int d^3\mathbf{x} \phi_i^*(\mathbf{x})
A^*(\mathbf{x})\psi(\mathbf{x})^2  \;\equiv\; (A_2)_i,\\
\sum_{pqj}a_j  c_p^* c_q^* \langle pq|ij\rangle 
&=&
\int d^3\mathbf{x}\psi^*(\mathbf{x})^{2} 
\phi_i(\mathbf{x}) A(\mathbf{x})  \;\equiv\; (A_2)_i^*,\\
\sum_j a_jV_{ij} &=&
\int d^3\mathbf{x} \phi_i^*(\mathbf{x}) \Vad(\mathbf{x})A(\mathbf{x})
\;\equiv\; (A_3)_i,\\
\sum_j a_jV_{ji} &=&
\int d^3\mathbf{x}A^*(\mathbf{x}) \Vad(\mathbf{x}) \phi_i(\mathbf{x})
\;\equiv\; (A_3)_i^*,
\label{eqn:A5}
\end{eqnarray}
where we have made use of the fact that $a_j = a_j^*$.  We have similar
expression for all the other auxiliary vector 
fields $b, f$, and $g$ and will refrain
from writing these all out.  We can write parts of 
Eqs.~(\ref{DQDQH2}), (\ref{DPDPH2}), (\ref{DQND2QH}) and (\ref{DPND2PH}) as
\begin{eqnarray}
\sum_{pqj}a_j \left[ 4 c^*_p c_q \langle pi|qj\rangle
+ c_p c_q \langle ij|pq\rangle   + c_p^* c_q^* \langle pq|ij\rangle \right]
&=& 4(A_1)_i + 2\mbox{Re}(A_2)_i,\\
\sum_{pqj}b_j \left[ 4c^*_p c_q  \langle pi|qj\rangle
- c_p c_q \langle ij |pq\rangle - c_p^* c_q^* \langle pq|ij\rangle\right ]
&=& 4(B_1)_i - 2 \mbox{Re}(B_2)_i,\\
\sum_{pqj}f_j \left[ 4 c^*_p c_q \langle pi|qj\rangle
+ c_p c_q \langle ij|pq\rangle   + c_p^* c_q^* \langle pq|ij\rangle \right]
&=& 4(F_1)_i + 2 \mbox{Re}(F_2)_i,\\
\sum_{pqj}g_j \left[ 4c^*_p c_q  \langle pi|qj\rangle
- c_p c_q \langle ij |pq\rangle - c_p^* c_q^* \langle pq|ij\rangle\right ]
&=& 4(G_1)_i - 2 \mbox{Re}(G_2)_i,
\end{eqnarray}
\begin{eqnarray}
\sum_j a_j[V_{ij} + V_{ji}] & = & 2\mbox{Re}(A_3)_i,\qquad
\sum_j b_j[V_{ij} + V_{ji}] \, = \, 2  \mbox{Re}(B_3)_i,\\
\sum_j f_j[V_{ij} + V_{ji}] & = & 2\mbox{Re}(F_3)_i,\qquad
\sum_j g_j[V_{ij} + V_{ji}] \, = \, 2  \mbox{Re}(G_3)_i.
\end{eqnarray}
The remaining terms are diagonal second derivatives of the form
 $\D^2 H = \sum_i(\D^2 H)_{ii}$, in which we have
matrix elements
\begin{eqnarray}
\sum_{pq} c^*_p c_q \langle pi|qi\rangle
&=&\int d^3\mathbf{x}|\phi_i(\mathbf{x})|^2|\psi(\mathbf{x})|^2 
   \;\equiv\; (M_1)_i,\\
\sum_{pq} c_p c_q \langle ii|pq\rangle 
&=&\int d^3\mathbf{x}\phi_i^*(\mathbf{x})^2 \psi(\mathbf{x})^2 
   \;\equiv\; (M_2)_i,\\
\sum_{pq}c_p^* c_q^* \langle pq|ii\rangle 
&=&\int d^3\mathbf{x}\psi^*(\mathbf{x})\phi_i(\mathbf{x})^2  
\;\equiv\; (M_2)^*_i,
\end{eqnarray}
and thus the appropriate parts of Eqs.~(\ref{DQH2}) and (\ref{DPH2}) are
\begin{eqnarray}
\sum_{pq} \left[ 4 c^*_p c_q \langle pi|qi\rangle
+ c_p c_q \langle ii|pq\rangle   + c_p^* c_q^* \langle pq|ii\rangle \right]
&=&
4(M_1)_i + 2\mbox{Re}(M_2)_i,\\
\sum_{pq} \left[ 4 c^*_p c_q \langle pi|qi\rangle
- c_p c_q \langle ii|pq\rangle   - c_p^* c_q^* \langle pq|ii\rangle \right]
&=&
4(M_1)_i - 2  \mbox{Re}(M_2)_i.
\end{eqnarray}
We have now explicitly outlined the calculation of all terms necessary to
calculate the temperature and chemical potential of the equilibrium dynamical evolution of a
microcanonical PGPE system.

\subsection{Summary}
Here we summarise all the components of the vector quantities required for the
evaluation of the temperature according to Eq.~(\ref{eqn:T}) making use of the definitions made in the
previous section for both choices of
vector operator $\D_Q$ and $\D_P$.  The inner products between these vectors 
found in Eq.~(\ref{eqn:T}) can be easily evaluated numerically.  In addition,
these are all the terms required for calculating the chemical potential as well.
\begin{eqnarray}
(\D_Q H)_i &=& \epsilon_i^2 Q_i 
+ \sqrt{ 2\epsilon_i}\mbox{Re}(W_i)
+ \Cnl \sqrt{2 \epsilon_i} \mbox{Re}(K_i),
\label{eqn:first}\\
(\D_P H)_i &=& P_i 
+ \frac{2}{\sqrt{ \epsilon_i}} \mbox{Im}(W_i)\
+ \Cnl\sqrt{\frac{2}{\epsilon_i}}\mbox{Im}(K_i),
\end{eqnarray}
\begin{eqnarray}
(\D_Q^2 H)_{ii} &=&  \epsilon_i^2 + 
\epsilon_i V_{ii}
+ \Cnl \epsilon_i[ 2(M_1)_i + \mbox{Re}(M_2)_i],\\
(\D_P^2 H)_{ii} &=& 1 + 
\frac{V_{ii}}{\epsilon_i }
+ \frac{\Cnl}{\epsilon_i }[2(M_1)_i -  \mbox{Re}(M_2)_i].
\end{eqnarray}

\begin{eqnarray}
\D_Q (|\D_Q N|^2)_i &=& 2\epsilon_i^2 Q_i,\\
\D_P (|\D_P N|^2)_i &=& 2\frac{ P_i}{\epsilon_i^2},\\
\sum_j (\D_Q H)_j (\D_Q^2 N)_{ij} &=& \epsilon_i (\D_Q H)_i,\\
\sum_j (\D_P H)_j (\D_P^2 N)_{ij} &=&\frac{(\D_P H)_i}{\epsilon_i}.
\end{eqnarray}

\begin{eqnarray}
\D_Q (|\D_Q H|^2)_i &=& 
2 a_i \epsilon_i^{3/2} + 
2 \sqrt{\epsilon_i}\mbox{Re}(A_3)_i
+ \Cnl \sqrt{\epsilon_i }[4(A_1)_i + 2\mbox{Re}(A_2)_i],\\
\D_P (|\D_P H|^2)_i &=& 
2\sqrt{\epsilon_i}b_i + \frac{2}{\sqrt{\epsilon_i}}
 \mbox{Re}(B_3)_i
+ \frac{\Cnl}{\sqrt{\epsilon_i}}
[4(B_1)_i - 2 \mbox{Re}(B_2)_i],
\end{eqnarray}

\begin{eqnarray}
\sum_j (\D_Q N)_j (\D_Q^2 H)_{ij}
&=& f_i \epsilon_i^{3/2} + 
\sqrt{\epsilon_i}\mbox{Re}(F_3)_i
+ \Cnl \sqrt{\epsilon_i }[2(F_1)_i +  \mbox{Re}(F_2)_i],
\\
\sum_j (\D_P N)_j (\D_P^2 H)_{ij}
&=&\sqrt{\epsilon_i}g_i + \frac{1}{ \sqrt{\epsilon_i}}
\mbox{Re}(G_3)_i
+ \frac{\Cnl}{\sqrt{\epsilon_i}}
[ 2(G_1)_i -  \mbox{Re}(G_2)_i].
\label{eqn:last}
 \end{eqnarray}

\section{Numerical results} 

In this section we outline the algorithm for applying the formulae of
the previous section to simulation results, and give examples of the results
obtained for the PGPE for trapped Bose gas systems.  

We have previously demonstrated in \cite{Davis2001b,Davis2002a,Blakie2005a}
that the PGPE will evolve any randomised initial field function to an
equilibrium that can be characterised by a temperature, and have made use of
this in the beyond mean-field calculation of the critical temperature of a
trapped Bose gas in \cite{Davis2005a}.

The basic requirement to determine the temperature in such calculations  is to
evaluate Eq.~(\ref{eqn:T}) over a sufficiently long period of simulation time
for the ergodic averaging to be effective.  Our procedure is as follows. 
Beginning with an initial random field configuration with a predetermined
energy, we run our simulation for sufficiently long such that we can be
reasonably sure that equilibrium has been reached (this can be checked as
described below).  This is typically for one hundred or more trap periods for
simulations of Bose gases in a harmonic potential.  We usually save one
thousand or more field configurations at equally spaced times throughout the
evolution, which are sufficiently separated in time  that there is little
correlation between samples.  

The temperature analysis is performed after the simulation is completed.  The
field configurations are stored as vectors of basis coefficients, and we make
use of efficient computational routines that avoid numerical aliasing  to
transform to and from a real space representation. For each saved field
configuration we calculate the vector components $P_i$ and $Q_i$, the
derivatives in Eqs.~(\ref{eqn:list1}--\ref{eqn:list2}), the auxillary fields
given by  Eqs.~(\ref{eqn:aux1},\ref{eqn:aux2}), and then their overlaps with
the various functions of the field $\psi(\mathbf{x})$ as in 
Eqs.~(\ref{eqn:A1}--\ref{eqn:A5}).  This allows us to calculate the vector
quantities defined in Eqs.(\ref{eqn:first}--\ref{eqn:last}) that appear in
Eq.~(\ref{eqn:T}). We then form the appropriate dot products of these
quantities to give a single sample of $\D_Q \cdot \mathbf{X}_T$ and  $\D_P
\cdot \mathbf{X}_T$.

These values are calculated for every saved field configuration.  These can
then be plotted versus time, and any initial transients before the field has
settled into equilibrium can be identified  (typically this a very small
fraction of the total simulation time.)  The quantities $\D_Q \cdot \mathbf{X}_T$
and  $\D_P \cdot \mathbf{X}_T$ are numerically distinct from one another before
equilibrium is reached, however they can be quite noisy and so sometimes it is
difficult to tell when they are in agreement.  We exclude the transients from
the final averaging to determine the temperature --- typically we discard at
least the first quarter of the time evolution to ensure accuracy.

\begin{figure}
\includegraphics[width=12cm]{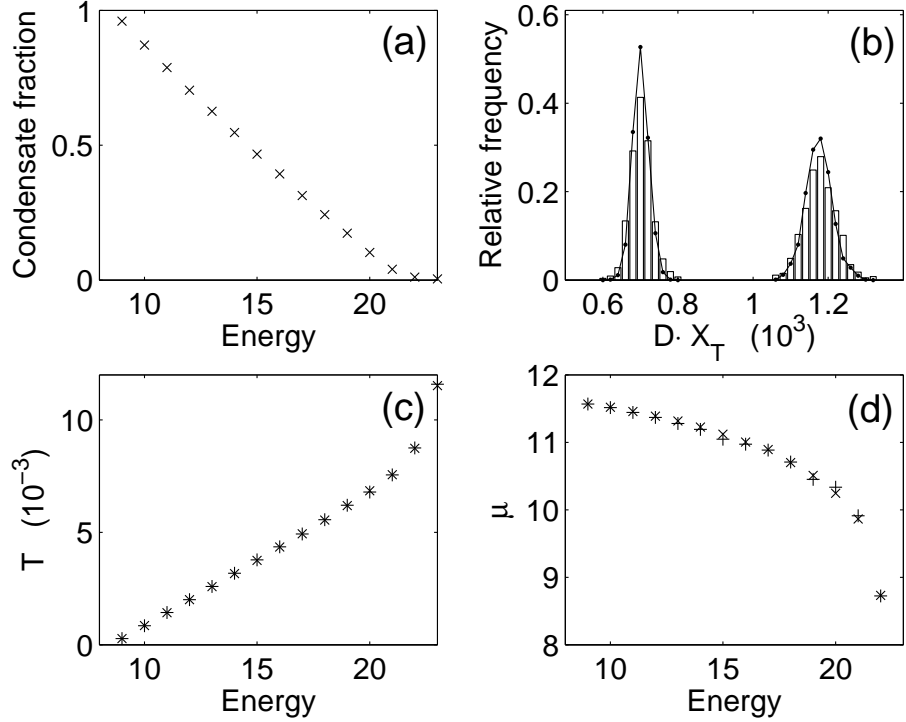}
\caption{Data extracted from PGPE simulations in equilibrium.    (a)  Condensate
fraction versus energy of simulation.  (b)  Histograms of the quantity 
$\D \cdot \mathbf{X}_T$ on the RHS of Eq.~(\ref{eqn:temp_eqn}) for the simulations
with $E = 10$ and $E = 11$.  The bars are for the operator $\D_Q$, and the 
dots connected by solid lines
for $\D_P$.  (c) Temperature versus energy.  Plusses: $T_Q$, crosses: $T_P$.
 (d) Chemical potential versus energy.  Plusses: $\mu_Q$, crosses: $\mu_P$.  All
 quantities are in dimensionless units as defined in the text.
}
\label{fig1}
\end{figure}

We now illustrate with a set of typical numerical results from
applying the method described in this paper to the PGPE for trapped Bose gas
systems.  We solve the PGPE (\ref{eq:GPE1}) for a harmonic trapping potential
$V(\mathbf{x}) = (x^2 + y^2 + 8z^2) /4$ and no additional potential
$\Vad(\mathbf{x})$.  The unit of length is $L = (\hbar / m \omega_x)^{1/2}$ and
energy $\epsilon_L = \hbar \omega_x$. We have a dimensionless energy cutoff of
$E_{\rm cut}=31$ such that there are 1739 classical modes, and the
data shown is for $\Cnl = 2000$.  We begin individual simulations with a
randomised initial condition with fixed energy, and evolve in dimensionless
time for until $\tau = 1200$, which is slightly more than 190 radial trap
periods.  We find that these simulations equilibrate very quickly, and use 1000
field samples over the last two-thirds of the evolution for the ergodic
averaging.  The results are shown in Fig.~\ref{fig1}, some of which have
previously been reported in Ref.~\cite{Blakie2005a}.

 Figure~\ref{fig1}(a) shows the
condensate fraction versus energy for the PGPE system (we remind the reader that
energy given by Eq.~(\ref{eqn:Hbasis}) is a conserved quantity in these calculations).  The condensate fraction
is determined by the Penrose-Onsager criterion: it is the largest eigenvalue
found by diagonalizing the single-particle density matrix formed by ergodic
averaging as described in \cite{Blakie2005a}.  It looks somewhat different to
the result for the full three-dimensional system due to the basis cutoff that is
present in these simulations.
Figure~\ref{fig1}(b) shows the distribution of the function
$\D \cdot \mathbf{X}_T$ for two different simulations, using  the
formulae for both $\D_Q$ and $\D_P$.  The temperature is found by the inverse of
the mean value of these distributions, and is shown for all simulations in
Fig.~\ref{fig1}(c).  We can see that the results of both calculations agree very
well, and so we can be confident that firstly the simulations have reached
equilibrium, and secondly that the numerical implementation of these
calculations are free from errors.  Figure~\ref{fig1}(d) shows the similar calculation for the
chemical potential.

One interesting point is that the width of the distributions  in
Fig.~\ref{fig1}(b) calculated using the operator $\D_P$ are slightly narrower
than those for $\D_Q$.  While this is of no consequence conceptually,
in practice the narrower the distribution the fewer samples are required
for an estimate of the mean to a required accuracy.  Given the large amount of
freedom in the choice of the operator $\D$, it seems quite possible that for
particular problems that some choices will be better than others.  We have
found situations where one of these distributions is signficantly narrower than
the other.  This has also been pointed out by Rugh \cite{Rugh2001a}.  However,
with no a priori way to estimate the width of the distributions, finding the
most accurate method of determing the temperature is a matter of trial and
error.

\section{Conclusions} 

To investigate the effect of critical fluctuations in Bose gases using the PGPE
in the microcanonical ensemble it is necessary to have accurate methods of
determining the thermodynamic temperature and chemical potential.  In this paper
we have explicitly outlined a method of how to do so using the assumption of
ergodicity and the dynamical evolution of the PGPE.  The method could 
potentially be applied
to other nonlinear classical field Hamiltonians.

\providecommand{\newblock}{}

\end{document}